\documentclass[aps,pre,twocolumn,flosuperscriptaddress]{revtex4}

\usepackage[english]{babel}
\usepackage[utf8x]{inputenc}
\usepackage[T1]{fontenc}

\usepackage[caption=false]{subfig} 
\usepackage{ragged2e} 
\DeclareCaptionJustification{justified}{\justifying}

\usepackage{amsmath}
\usepackage{amssymb}

\usepackage{graphicx}
\usepackage[colorlinks=true, allcolors=blue]{hyperref}

\begin{document}
\title{Anomalous behavior of Replicator dynamics for the Prisoner's Dilemma on diluted lattices.}
\author{Fernanda R. Leivas}
\email{fernanda.leivas@ufrgs.br}
\affiliation{Instituto de F\'{\i}sica, Universidade Federal do Rio Grande do Sul CP 15051, 91501-970 Porto Alegre RS, Brazil}
\author{Heitor C. M. Fernandes}
\email{heitor.fernandes@ufrgs.br}
\affiliation{Instituto de F\'{\i}sica, Universidade Federal do Rio Grande do Sul CP 15051, 91501-970 Porto Alegre RS, Brazil}
\author{Mendeli H. Vainstein}
\email{vainstein@if.ufrgs.br}
\affiliation{Instituto de F\'{\i}sica, Universidade Federal do Rio Grande do Sul CP 15051, 91501-970 Porto Alegre RS, Brazil}

\begin{abstract}
In diluted lattices, cooperation is often enhanced at specific densities, particularly near the percolation threshold for stochastic updating rules. However, the Replicator rule, despite being probabilistic, does not follow this trend. We find that this anomalous behavior is caused by structures formed by holes and defectors, which prevent some agents from experiencing fluctuations, thereby restricting the free flow of information across the network. As a result, the system becomes trapped in a frozen state, though this can be disrupted by introducing perturbations. Finally, we provide a more quantitative analysis of the relationship between the percolation threshold and cooperation, tracking its development within clusters of varying sizes and demonstrating how the percolation threshold shapes the fundamental structures of the lattice.
\end{abstract}

\maketitle
\section{Introduction}
The emergence and maintenance of cooperation are crucial phenomena in nature, essential to the evolution of cells, multicellular organisms and  the organization of human society as a whole~\citep{Axelrod84,Nowak06,bowless11,Kerife02}. Cooperative patterns are observed not only between organisms of the same species~\cite{West2016} but also between those of different species~\cite{Bacucher85}. To understand the limits and nature of cooperation, computational and mathematical models, such as Evolutionary Game Theory, have been widely implemented and studied. Its origins lie in the study of mathematical models in behavioral sciences, particularly in economics~\cite{Neumann44}, and its success in predicting complex behavior from simple interactions led to its application in diverse fields such as evolutionary biology~\cite{Maynard82,Parker90,Weibull95,Dugatkin98,Cleve2024}, political science, psychology~\cite{Mattew93,Schalr2001,Perati15} and, more recently, in studying the spread of diseases and pandemic patterns~\cite{amaral2021a, Kabir2020, Martin2020,HAFEZALKOTOB2023, KULSUM2024114364, MENDONCA2024128749}. With the aim of studying and predicting behaviors among interacting individuals, whether economically or evolutionary, it uses basic concepts such as cooperation and competition. What drives individuals to adopt one or another behavior is a fundamental question that has yet to be fully answered.
  
The foundation of complex human societies may be linked to our capacity for large-scale cooperation, even when individuals do not receive the maximum personal benefit by cooperating~\cite{harari2015sapiens}. The Prisoner's Dilemma (PD) is a simple yet compelling game that helps us understand this type of human interaction, as it explores the persistence of cooperation even among selfish individuals who might be tempted to defect to maximize their personal gain. According to classical game theory, mutual defection (selfishness) in the PD represents the only Nash equilibrium~\cite{Nash50}. What makes this game particularly interesting is that when played repeatedly in a population with short-term memory over many rounds (as in Evolutionary Game Theory), it allows cooperative strategies to persist~\cite{AleHa81}. This phenomenon can be explained by the assumption that players are not fully rational or that rationality is not common knowledge in repeated social dilemmas~\cite{KREPS1982253,SzFa07}.

The spatial structure, as a characterization of the environment where the players are inserted, has a decisive impact on the maintenance of cooperation~\cite{NoMa92,NoMa93,Namaiw97,PoGoFlMo07, XuZAng16, Arenzon2018, Chen2018, Flores2022}. Previous studies show that cooperators are able to survive even with high temptations to defect by forming spatial compact clusters that protect the interior against invasion~\cite{DoHa05,PeSz08}. Inspired by the success of spatial reciprocity, many types of complex networks were studied in the context of game theory, such as scale-free networks,  growing populations, hierarchical structures~\cite{SaPa05, SaPaLe06, PoncGoFlo07, LozAr08, LeeHol11} and adaptive multilayer networks~\cite{Chen_2021}. Diluted lattices are useful as a model to understand the impact of heterogeneous environments on the interactions between individuals~\cite{VaAr01, 10.2307/2296617,Amaral2017}. The presence of vacant sites allows for movement, simulating migration, and studies have shown that allowing agents to move can enhance cooperation~\cite{VaSiAr07,  VaSiAr07a, Xiao2022, Xiao_2020, Amaral2021}. Also, understanding how cooperation is affected by varying system density can provide insights into animal behavior under crowded or sparse conditions.  Research indicates that social pathologies, such as increased stress, aggression, and even infanticide, can be exacerbated in densely populated environments~\cite{Calh62}.

Wang {\it et al.}~\cite{WaSzPe12,WaSzPe12b} demonstrated that percolation~\cite{RevModPhys.45.574} directly influences the outcome of evolutionary simulations, showing that a peak in cooperative behavior is expected near the percolation threshold of the underlying lattice. Subsequent studies using different games and network models, such as multilayer populations~\cite{WANG2023113154}, also observed that the percolation threshold may favor cooperation. These studies highlight that the dynamics and distribution of individual clusters, as density varies, are crucial for the success of cooperation~\cite{Yang_2018}. At densities near the percolation threshold, agents are sufficiently connected to support large cooperative clusters while being diluted enough to prevent the invasion of defectors. It has also been demonstrated that this direct relationship is observed only in stochastic update rules, unlike deterministic dynamics, such as the  ``choosing the best'' rule~\cite{VaAr01}, which displays a strong dependence on initial conditions. Interestingly, off-lattice models have shown that the percolation of defector clusters is also critical for preventing their extinction~\cite{VaCaAr14}.

This work begins by analyzing the behavior of clusters of individuals of various sizes over time, using the finite Fermi transition rule in the context of pairwise interactions. This part of the research was initiated to develop a more accurate understanding of how percolation, by influencing the spatial distribution of agents, relates to cooperation. After establishing this foundation, we focus on the Replicator (REP) update rule, also within the framework of pairwise interactions. The Replicator dynamics, widely applied in evolutionary  games~\cite{hofbauer_sigmund_1998,SzFa07,Cressman2014,Mittal2020,Weitz2016AnOT,Cooney2019}, is  very important since it operates under the Darwinian assumption that the growth rate per capita depends on how well a strategy performs relative to the average population performance.

Surprisingly, compared to Fermi, we found that the REP transition rule exhibits unexpected behavior concerning the optimal population density. Our studies reveal that despite using probabilistic updates, deterministic patterns can emerge in REP due to the absence of irrationality (which will be defined in the next section) in this rule. In summary, our findings provide valuable insights into cooperation within stochastic dynamics and its behavior in diluted lattices.

\section{Model}

The model features a two-dimensional square lattice of size $ N=L^2$ with periodic boundary conditions. Each site on the lattice represents either an individual or a vacant spot (hole), distributed randomly. Individuals are assigned as cooperators ($C$) or defectors ($D$) initially with equal probability. The occupancy of each site depends on the density $\rho = (N_C+N_D)/N $, where $N_C$ and $N_D$ are the initial number of cooperators and defectors, respectively, and the total number of agents, $N_A=N_C+N_D$, is kept constant during the simulation. 

The players interact with others belonging to their von Neumann neighborhood (4 nearest neighbors) and sum all their resulting payoffs according to the Prisoner's dilemma game. Self-interaction is considered only when explicitly stated; in this case, besides interacting with their 4 nearest neighbors, they also interact with themselves, which leads to an enhancement in cooperation~\cite{NIU2018133} since only cooperators receive an additional reward. The payoff for each type of interaction is: in the case of mutual cooperation, both receive the reward $R$; and for mutual defection, the punishment $P$. A cooperator receives $S$ (sucker) if playing against a defector, while the defector receives the temptation $T$. The values should obey $T> R> P> S $ and $2R > T + S $ to characterize a Prisoner's dilemma: in a single encounter,  the individuals are always tempted to defect. To allow us to study a  unique parameter, we adopt the simplified re-scaled payoff matrix for the weak Prisoner's dilemma~\cite{NoMa92, Vainstein2014}: $ S = P = 0 $, $ R = 1 $, $ T = b $ ($ b>1 $). No payoff is received if interacting with a hole.

Updating rules can be deterministic, when individuals always adopt the strategy with the highest performance, or probabilistic, allowing individuals with a lower payoff to keep their strategy, and possibly also to adopt a strategy with lower benefits. We use stochastic pairwise comparison strategy update rules under synchronous update: first all individuals interact with their neighbors accumulating payoffs from all games. Subsequently, each player chooses a random neighbor and,  if it is not a hole, compares their respective payoffs. The decision to adopt or not the neighbor's strategy depends on one of the following transition probabilities $W$, kept fixed throughout the simulation unless otherwise stated. Therefore, an individual $x$ with payoff $P_x$ and strategy $s_x$ will adopt its neighbor's ($y$ with payoff $P_y$) strategy $s_y$, with the probability $W$ given by one of the following rules
\begin{itemize}
\item Fermi transition rule: 
\begin{equation*}
W(s_x \to s_y) = \frac{1}{1+e^{-(P_y-P_x)/K}},
\end{equation*}
where $K$ is a noise parameter that allows agents to maintain their strategy even if it has a worse performance, Moreover, it also allows irrational decisions-- i.e., a change to a strategy with worse performance~\cite{Szabo98}. We adopt a fixed value of $K=0.1$;

\item Replicator transition rule (REP)~\cite{OHTSUKI200686}:
\[ W(s_x \to s_y)  = \left\{ \begin{array}{ll}
         \frac{P_y-P_x}{4(T-S)}, & \mbox{for $P_y>P_x$}\\
        0, & \text{ otherwise}.\end{array} \right. \] 
        This rule is based on replicator dynamics for infinite populations~\cite{gintis2000game}, and under this rule, agents can retain their strategy even if it has the worst performance. However, unlike the previous rule, it does not permit irrational decisions. When $P_y > P_x$, the denominator is the maximum payoff difference; in the case of self-interaction, it becomes $(4(T-S)-1)$ due to the additional cooperation reward.
 \end{itemize}
 
Each Monte Carlo step (MCS) allows every player to adopt the strategy of one of its neighbors, depending on the probabilities above. Simulations were carried out with $L=512$  and the typical relaxation times needed to achieve a stationary state varied from $10^4$  to $10^6$ MCS. Lattice dilution and mobility can substantially increase the time taken to reach a stationary state~\cite{VaSiAr07a}. All our results are averages taken during the stationary period. 

For faster results, we used a CUDA simulation. Each site is described by a state variable: 1 for a cooperator (C), 2 for a defector (D), and 0 for an empty site. The chosen von Neumann spatial structure allows simple algorithms to take full advantage of the GPU architecture. Initially, we employed a straightforward approach where the system is loaded into global memory, and operations are executed directly from there without utilizing shared memory, serving as a benchmark for more complex implementations. During the MCS, one GPU kernel calculates and stores the payoffs of all lattice sites in global memory before launching the GPU kernel responsible for updating strategies. By knowing all payoffs, it is possible to update all strategies synchronously until asymptotic results are reached.  In order to obtain a satisfactory result, an average of  10 different samples was performed.
    
\section{Results}
\subsection{Fermi update rule}

\begin{figure}
\includegraphics[angle=270, width=1\columnwidth]{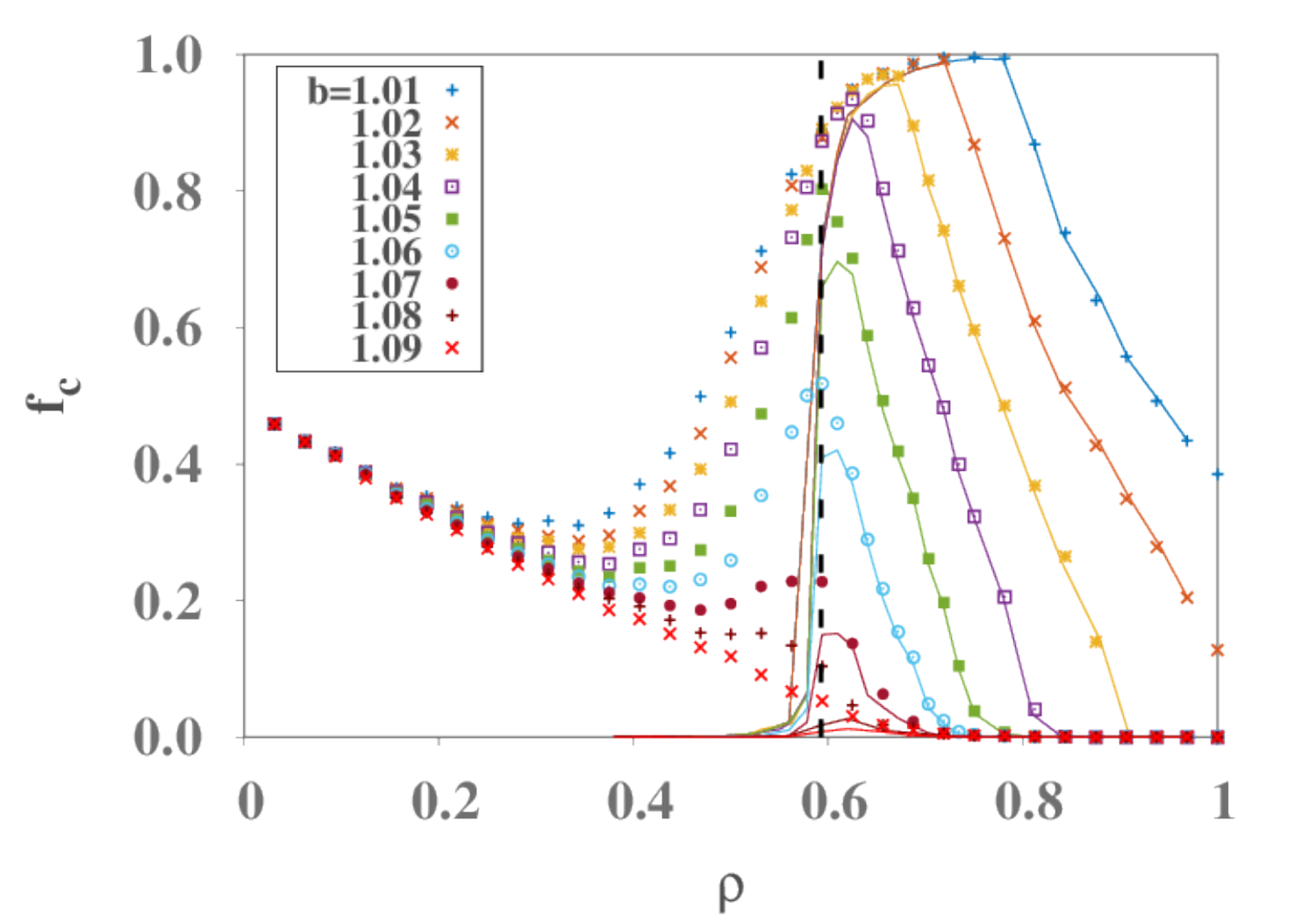}
\caption{The asymptotic fraction of cooperators $f_c$, as a function of the population density $\rho$, has an optimal value whose location approaches the site percolation threshold of the square lattice (vertical dashed line, $\rho_p\approx0.59$) as the temptation, $b$, increases towards the value that leads cooperation to extinction in the Prisoner's Dilemma under Fermi dynamics. The points represent the global population fraction, while the lines represent the fraction of cooperators in the largest cluster, $N_{cb}/N_A $. Note that lines overlap with the points only for values $\rho>0.59$.}
\label{fermi}
\end{figure}

\begin{figure}
\centering
    \subfloat(a){{\includegraphics[angle = 270, width = 8.0cm]{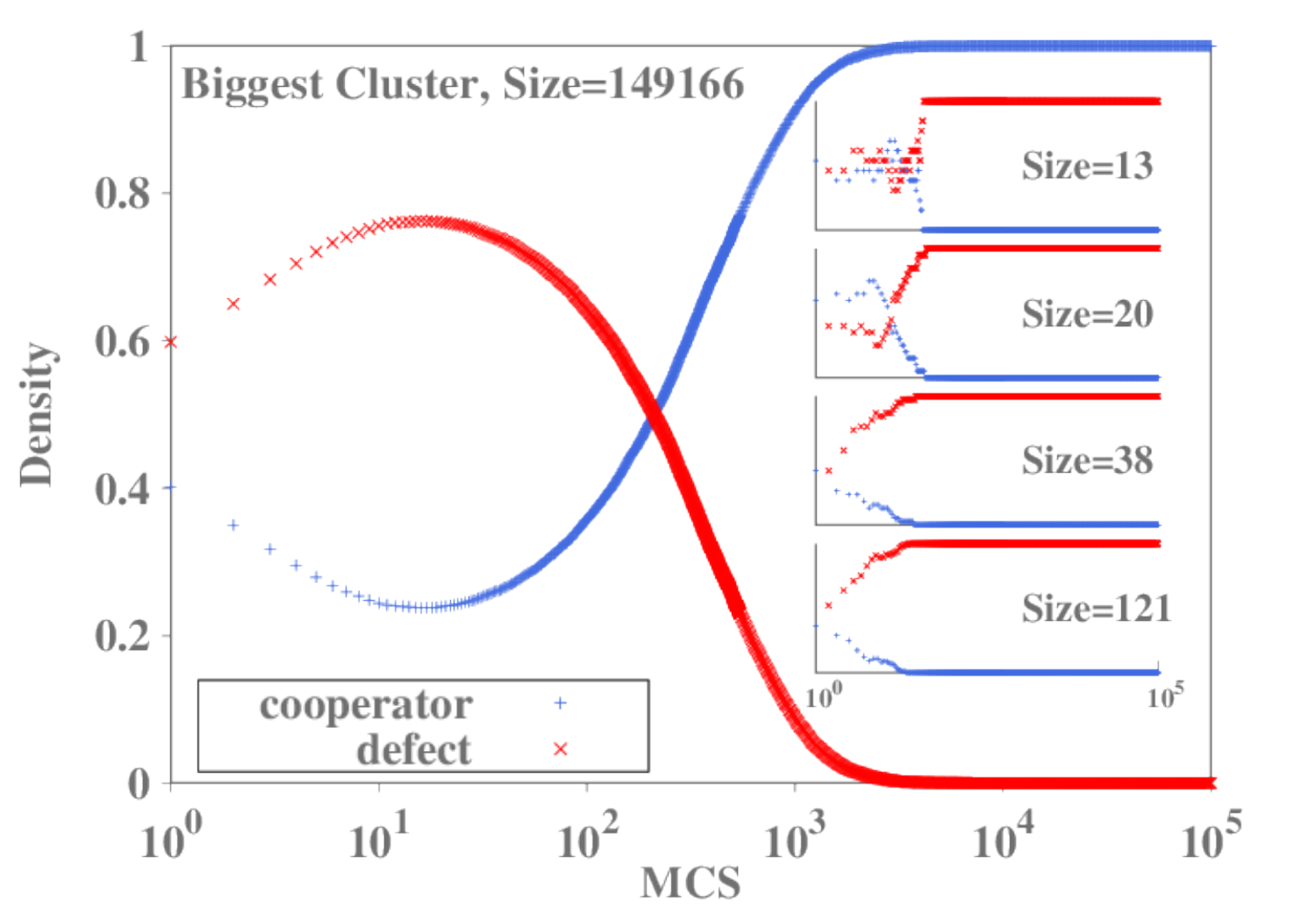} }}%
    \subfloat(b){{\includegraphics[angle = 270, width = 8.0cm]{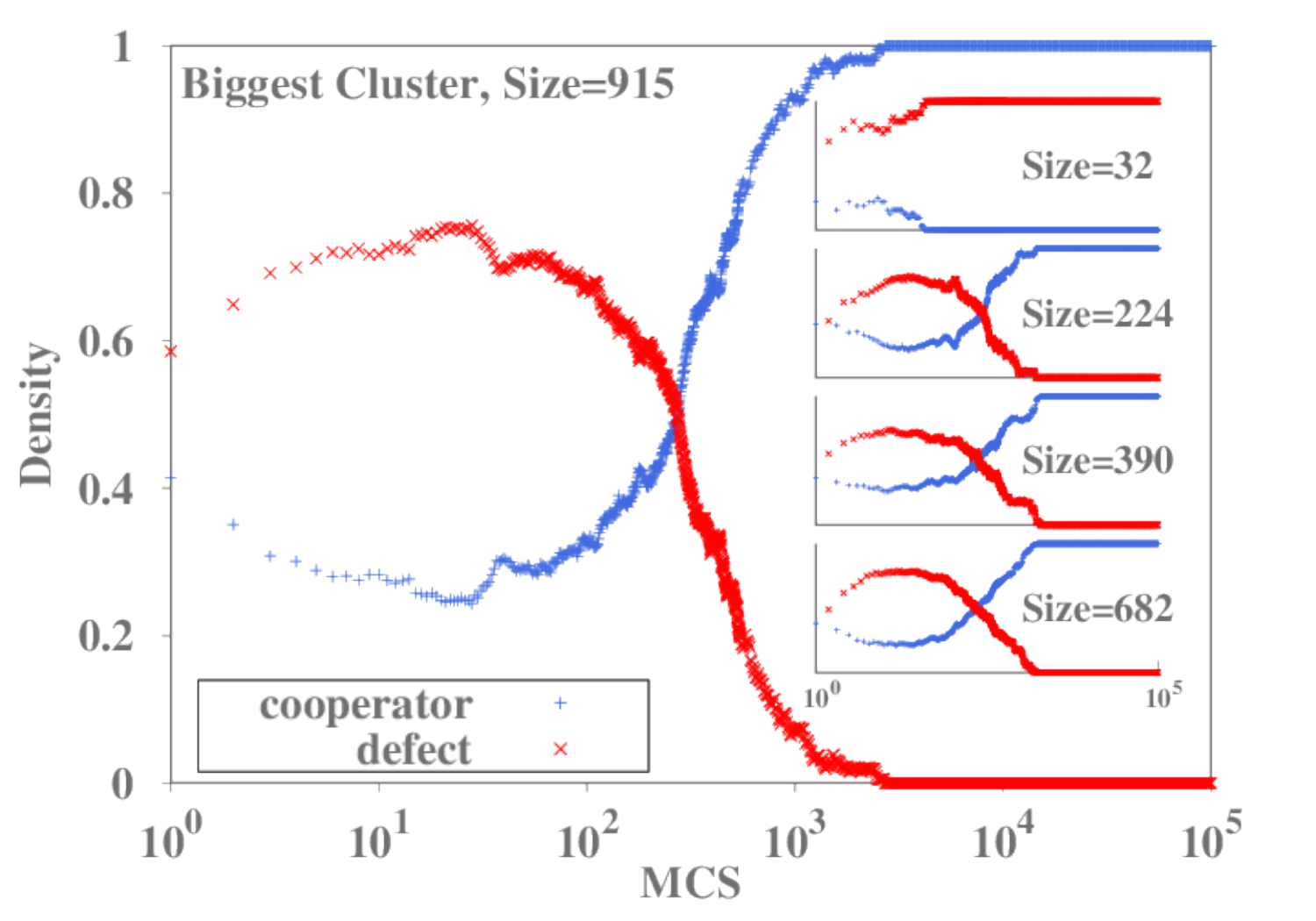}  }}%
\caption{The temporal evolution of strategies within clusters depends on their sizes. Blue points represent cooperators, while red points represent defectors. (a) Evolution in the largest cluster, comprising $92\%$ of the individuals in the simulation at a lattice density of $\rho=0.62$ (above the percolation threshold). The insets display the same data for smaller clusters of various sizes. (b) Evolution of strategies within the largest cluster, which comprises $0.5\%$ of the individuals in the simulation at a lattice density of $\rho=0.53$ (below the percolation threshold). In both graphs, the temptation to defect is fixed at $b=1.01$. This data was obtained from a single sample, providing reliable statistics due to the large lattice size.}
\label{clust_temp} 
\end{figure}

\begin{table*}[ht!]
\caption{Percentage of clusters dominated by cooperators, defectors, or comprised of both strategies for large non-percolating clusters (size $>100$) and small clusters ($2 <$ size $ \leq  100$). These data represent the cumulative results from all clusters across the densities studied in this work.} \label{big_clus_table}
\begin{tabular}{|p{0.8cm}||p{1.25cm}|p{1.25cm}|p{1.25cm}|p{1.25cm}|p{1.25cm}|p{1.25cm}|}
\hline
 \multicolumn{7}{|c|}{\hspace{0.8cm}2<cluster size$\leq$100 \hspace{1.5cm} 100<cluster size}\\
\hline
  b &Coope.  &Defec. &Mixed  &Coope.  &Defec. &Mixed \\
 \hline
 1.01 & 34.05\% & 65.95\% & 00.00\% & 98.36\% & 01.64\% & 00.00\% \\
 1.02 & 30.30\% & 69.70\% & 00.00\% & 96.16\% & 03.84\% & 00.00\% \\
 1.03 & 26.83\% & 73.17\% & 00.00\% & 94.87\% & 05.13\% & 00.00\% \\
 1.04 & 23.12\% & 76.87\% & 00.00\% & 86.45\% & 13.05\% & 00.50\% \\
 1.05 & 19.95\% & 80.05\% & 00.00\% & 62.25\% & 22.12\% & 15.63\% \\
 1.06 & 17.35\% & 82.65\% & 00.00\% & 25.63\% & 43.12\% & 31.25\% \\
 \hline
\end{tabular}
\label{table_cluster}
\end{table*}

Before presenting the anomalous behavior, we first examine the expected outcome and its underlying causes by analyzing the stochastic Fermi update rule. It is well-established that the site percolation threshold, $\rho_p$, of a lattice significantly influences cooperation in systems under Fermi dynamics and a stochastic version of the ``choosing the best'' strategy. Specifically, the maximum fraction of cooperation tends to occur as $\rho$ approaches the site percolation threshold at values of $T$ that nearly drive cooperation to extinction~\cite{WaSzPe12}, as illustrated in Fig.~\ref{fermi}. However, to our knowledge, no quantitative measures have been provided to make this relationship more intuitive. To gain a deeper understanding of this phenomenon, we explore the evolution of clusters under the Fermi update rule, examining how factors such as isolation, inter-group connections, and variations in density and size influence cooperation.

The Hoshen-Kopelman~\cite{HoKo76} algorithm was employed to identify player clusters and determine the fraction of cooperators and defectors in the asymptotic state, as well as to analyze their evolution over time. Figure~\ref{clust_temp} illustrates the evolution of strategies for the largest cluster (with temptation $ b = 1.01 $). The insets show examples of how smaller clusters evolve at the same density. The densities were chosen to be just above, (a) $\rho=0.62$, and below, (b) $\rho=0.53$, the site percolation threshold ($\rho_p\approx0.59$, for an infinitely large square lattice~\cite{Newman2000,Oliveira2003}). In Fig.~\ref{clust_temp}(a), the system mostly consists of a large percolating cluster containing $ 92 \% $  of the individuals in the simulation, along with several smaller clusters. In contrast,  the largest cluster contains less than $ 1 \% $ of individuals in Fig.~\ref{clust_temp}(b). In the former case, the dynamics of the system is dominated by the percolating cluster. In the latter case, it results from the combined effects of many isolated, independently evolving clusters. This difference can be seen in Fig.~\ref{fermi}, where the continuous lines represent the fraction of cooperators in the largest cluster as a function of $\rho$, whereas the points represent the fraction of cooperators in the whole population. For $ \rho > \rho_p $, the points are superimposed on the lines since the system is basically comprised of the largest cluster;  for $ \rho <\rho_p $, the lines quickly  tend to zero because even the largest clusters represent a small percentage of the system. In the insets of Fig.~\ref{clust_temp}(a) and (b),  we see that defectors tend to dominate smaller clusters (sizes = $13$, $17$, $20$, $32$, $38$). However, there is a threshold cluster size beyond which cooperators can resist the initial defector invasion, fostering an environment where cooperation can thrive (sizes = $134$, $224$, $390$, $682$). Table~\ref{table_cluster} shows the dominance of defectors in small clusters and the success of cooperation in larger, non-percolating clusters.
    
The study of lattice behavior, focusing on the evolution of clusters, allows us to understand the global asymptotic outcome of the strategies; for  low  densities $0<\rho<0.3$, individuals become more and more isolated and  $f_c \to 0.5$, the initial condition,  when $\rho \to 0$. For intermediate densities, $ 0.3 <\rho <0.59 \approx \rho_p $, the closer we are to $ \rho_p $, the greater is the number of clusters large enough to support cooperation, and the higher the cooperation fraction. However, when the density becomes greater than the percolation threshold $ \rho> \rho_p $, and the temptation is low $ 1.01 < b <1.05 $, lattice connectivity will favor cooperation only to a certain extent, notably, until the peak appears. In the high density regime, the number of defectors around cooperative clusters will be large enough to disrupt them, reducing the number of cooperators and weakening cooperation as a whole. For these cases the cooperative peak emerges at higher densities than $\rho_p$. However, if temptation is high enough ($b>1.05$), lattice connectivity will favor defectors immediately when $ \rho> \rho_p$ and cooperation will vanish. For this reason, the relationship between $ \rho_p $ and the peak of cooperation is more evident for high temptations, when cooperation is close to extinction.

\subsection{Replicator update rule}
\begin{figure}
\centering
\includegraphics[angle=270,width=1\columnwidth]{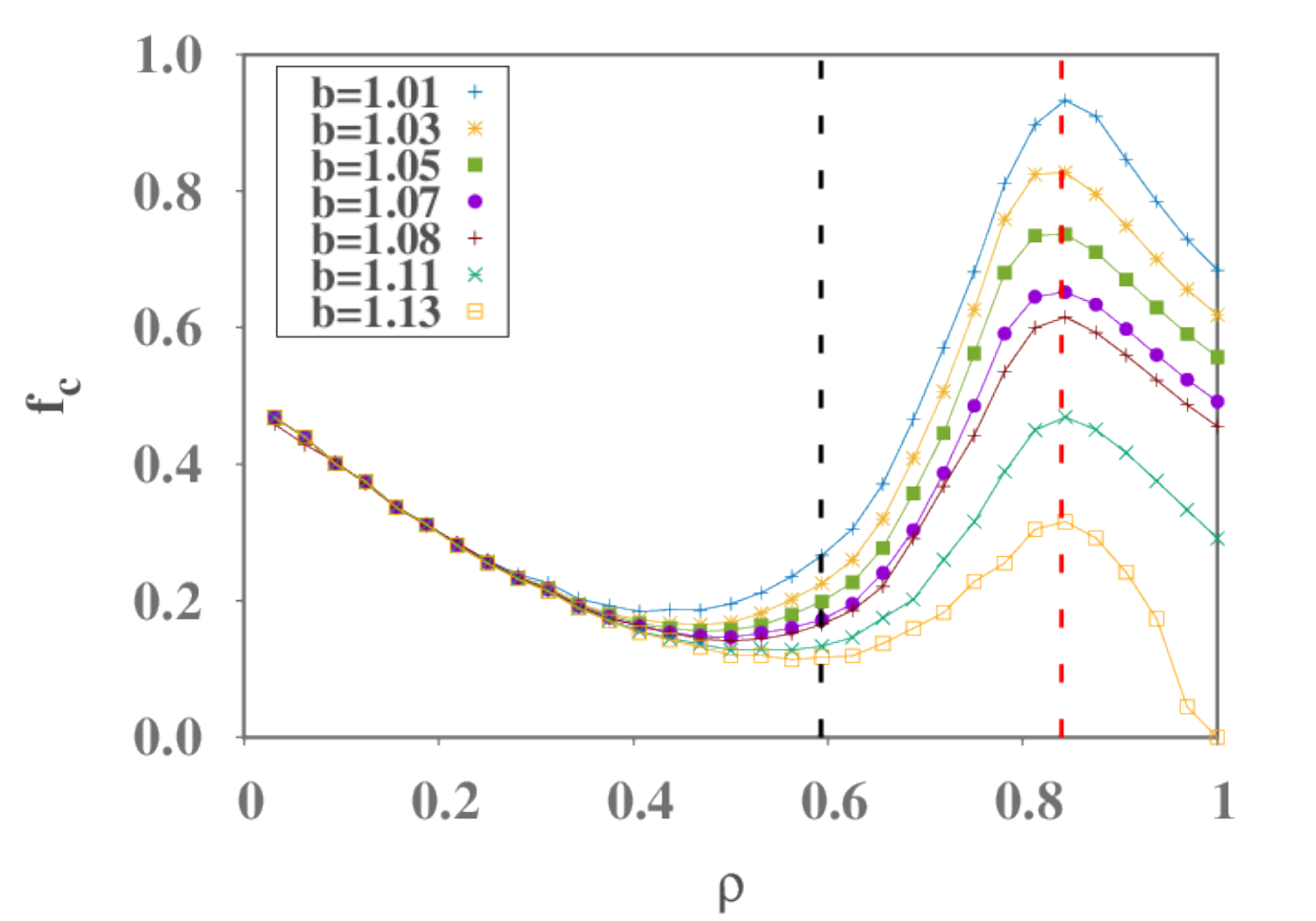}
\caption{The asymptotic fraction of cooperators $f_c$, as a function of population density $\rho$, also has an optimal value for Replicator dynamics at different levels of temptation to defect, $b$, in the Prisoner's Dilemma on the square lattice with synchronous updating under the Replicator update rule. However, unlike Fermi dynamics, its location does not coincide with the site percolation threshold, $\rho_p\approx 0.59$ (vertical black dashed line). The red dashed line pinpoints the position of the maximum in cooperation at $\rho^*\approx 0.85$.}
\label{rep}
\end{figure}

The location of the cooperative peak, $\rho^*$, as shown in Fig.~\ref{rep}, does not appear to be related to $\rho_p$, unlike for the Fermi rule (Fig.~\ref{fermi}). Instead, for all values of $b$, $\rho^* \approx 0.85$ when using the replicator update rule (REP).

\begin{figure}
\centering
\includegraphics[angle=270,width=1.0\columnwidth]{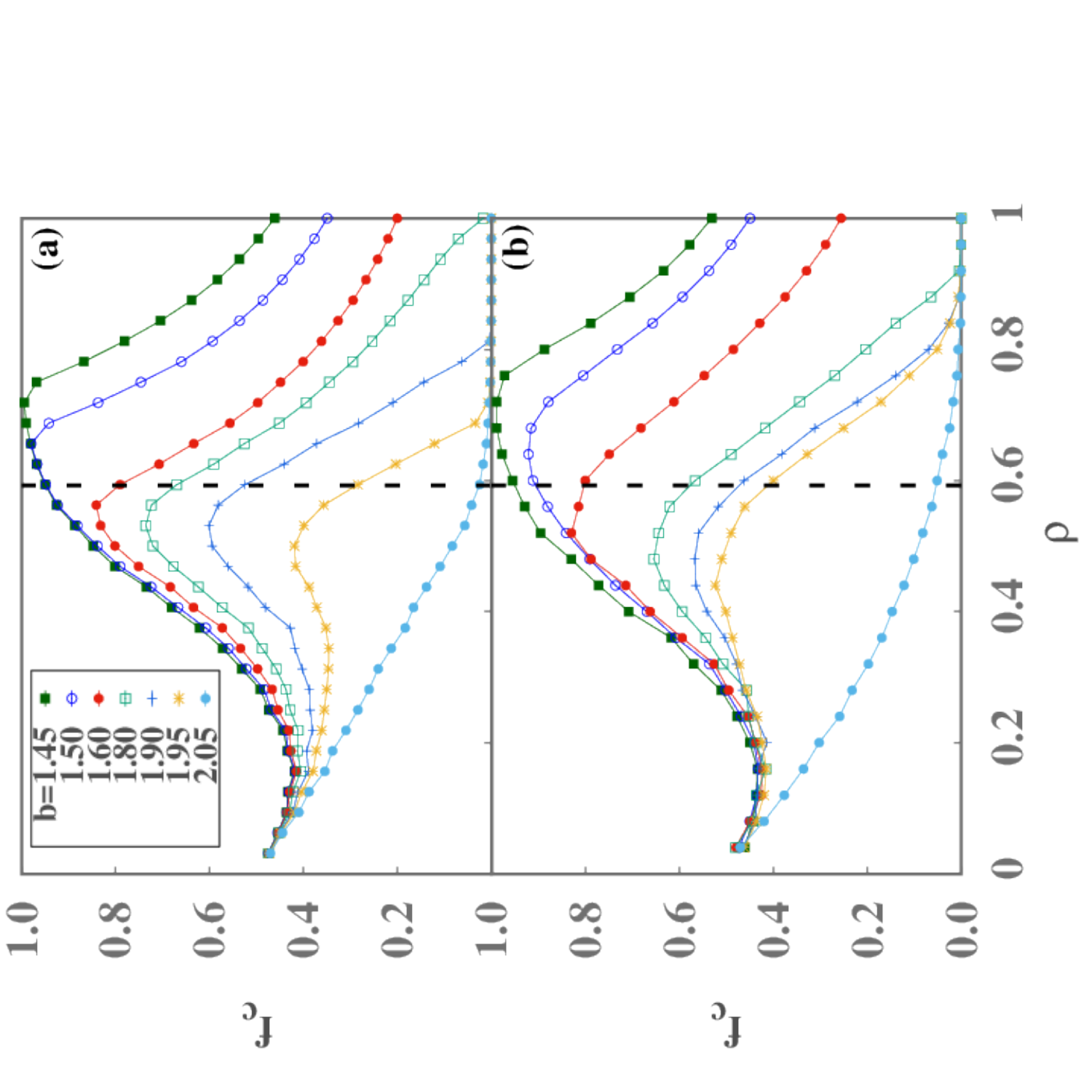} 
\caption{The asymptotic fraction of cooperators $f_c$, as a function of population density $\rho$, for the Prisoner's Dilemma on the square lattice with synchronous updating under both (a) Fermi and (b) Replicator dynamics, now with self-interaction,  have an optimal value of cooperation whose location is again near the 
site percolation threshold, $\rho_p\approx 0.59$, represented by the vertical black dashed line.}
\label{densc1}
\end{figure}

In order to better understand the reason for this divergence, we will see what happens in a similar system differing only by the introduction of self-interaction, a common model variation~\cite{NIU2018133}. The outcome of this modification is shown in Figs.~\ref{densc1} (a) and (b)  which  again presents $f_c$ as a function of $\rho$ for Fermi and REP update rules, respectively. The remarkable fact here is that both transition rules display a qualitatively similar behavior, specially when we compare the cooperative fraction peak $\rho^*$, contrary to what was found in the case without self-interaction. It is also relevant to note that, in this scenario, the optimum population value $\rho^*$ for cooperation decreases, moving away from the percolation threshold. This is not  unexpected if we bear in mind that self-interaction favors only cooperators,  allowing  them  to survive even  in small clusters at lower densities. Therefore, in the case of self-interaction, the REP update rule behaves just as expected from a stochastic transition rule.

\begin{figure} 
\centering
\includegraphics[angle=270,width=8.5cm]{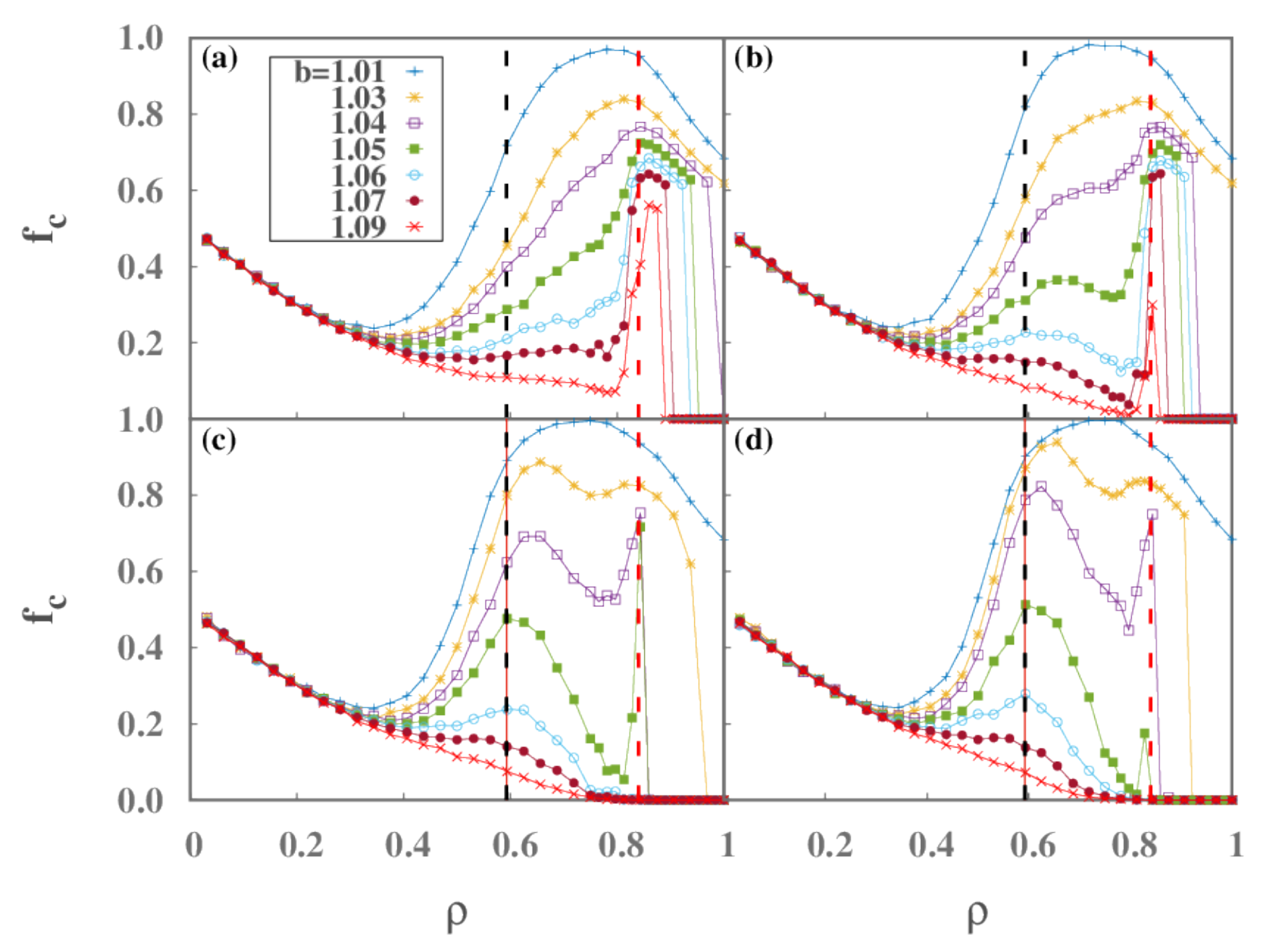}
\caption{The asymptotic fraction of cooperators $f_c$, as a function of the population density $\rho$, is shown for different values of the temptation to defect $b$, in the Prisoner's Dilemma on the square lattice under synchronous updating. First, the simulation is performed with the Replicator update rule until asymptotic values of $f_c$ are reached. Then, for a short period,  the dynamics is  switched to the Fermi update rule before returning to the Replicator rule. The amount of time spent with the Fermi dynamics is given as a fraction of the total number of Monte Carlo steps  (a) $\phi=0.5\%$, (b) $\phi=1\%$, (c) $\phi=5\%$ and (d) $\phi=10\%$. The black dashed vertical lines correspond to the site percolation $\rho_p\approx 0.59$ and the red dashed lines pinpoint the position of the maximum in cooperation $\rho^*\approx 0.85$ in the unperturbed Replicator dynamics.}
\label{perturb}
\end{figure}   

\begin{figure*}
\includegraphics[angle = 0, width = 15.5cm]{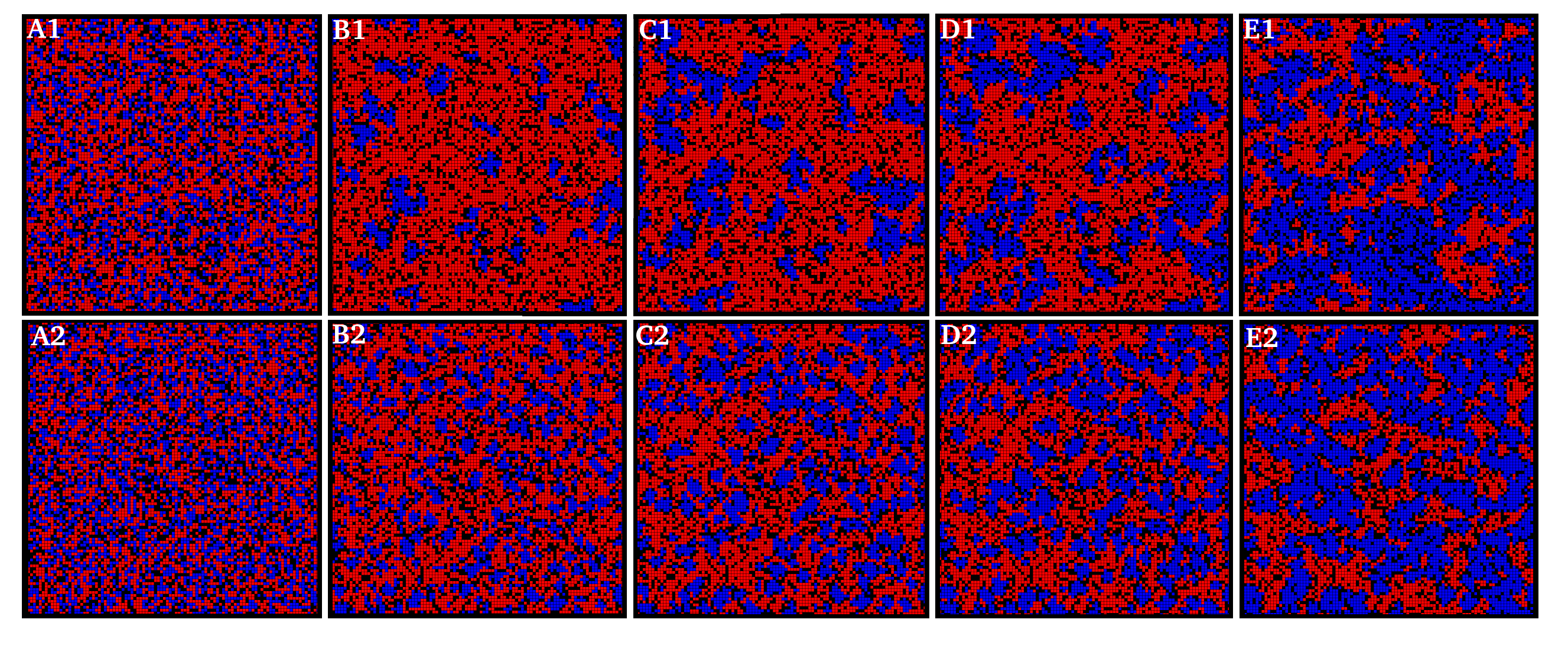} 
\caption{Snapshots of the temporal evolution under the Fermi (top row-- A1, B1, C1, D1, E1) 
 and the Replicator (bottom row-- A2, B2, C2, D2, E2) rules in a square lattice with $ b = 1.03 $, $ \rho = 0.75 $ and $ L = 100 $. Cooperators are blue (dark gray), defectors are red (light gray) and holes are black. Both cases converge to approximately the same cooperation density, despite having different relaxation times and structural features. Time in Monte Carlo Steps increases from left to right: $t_A=0$, $t_B=100$, $t_C= 300$, $t_D=500$, and $t_E=800$. 
}
\label{snap_rep_fermi}
\end{figure*}

\subsubsection{The perturbed Replicator rule}
To investigate the stability of the steady-state reached by the system under REP dynamics without self-interaction, we analyze how it responds to a perturbation  caused by temporarily changing its dynamics. The system is first allowed to reach the asymptotic state using the REP update rule; then, the dynamics is switched to Fermi's rule for a short period before returning to the original rule. Finally, the system is allowed to reach its new steady state before measurements of the asymptotic values of cooperation are made. The duration of the perturbation period was varied, resulting in different intensities, characterized by
\begin{equation}
    \phi=\frac{MCS_{F}}{MSC_{T}},
\end{equation}
defined as the ratio of the number of Monte Carlo steps using the Fermi update ($MCS_F$), and the total number of Monte Carlo steps ($MCS_{T}$). Figure~\ref{perturb} shows the outcome for $\phi=0.5\%$, $1\%$, $5\%$, and $10\%$ and reveals an interesting consequence of the perturbation: as we increase the value of $\phi$, it becomes clear that there exist two cooperative peaks, or their superposition, at low values of $b$. Comparing the outcome with the case without perturbation (Fig.~\ref{rep}), for $\rho > 0.85$, the asymptotic fraction of cooperators reaches the same value, provided they are not driven to extinction during the process, indicating that this is a stable region under perturbation. However, for lower densities, the curves do not return to their original state, and another peak appears. Thus, the dynamics for these values of $\rho<0.85$ could be considered metastable.  We have also performed the opposite procedure, perturbing the Fermi dynamics with REP; however,  the outcome is exactly the same as in its original version (for this reason, we do not show the data), indicating that the Fermi dynamics is stable, as expected.

\subsubsection{Temporal evolution and persistence}

It is useful to analyze and compare the temporal evolution of the system under the two transition rules to uncover the mechanisms responsible for the anomalous behavior of the REP dynamics. Figure~\ref{snap_rep_fermi} shows snapshots of the system for the two dynamics: the top row (A1, B1, C1, D1, E1) refers to   Fermi's rule, and the bottom row (A2, B2, C2, D2, E2), to REP. The times at which they were taken vary from $0$ MCS to $800$ MCS, with $ b = 1.03 $ and $ \rho = 0.75 $. This value of $ \rho $ was chosen, because in this case both dynamics reach the same value of $ f_c \approx 0.60 $.  Although the systems reach similar values of $f_c$ and appear quite similar near convergence (E1 and E2), the temporal evolution of the distribution of strategies differs between them. In the early stages (fewer than $100$ MCS), there is a significant shift in the distribution of strategies under the Fermi rule (B1) compared to the initial state (A1). However, under the REP rule, patterns from the initial condition (A2) can still be observed and persist in subsequent steps. Clusters of defectors, which form large low payoff regions, remain fixed throughout the simulation; similarly, clusters of cooperators, which create high-payoff areas, are also sustained. Agents within both types of clusters maintain their strategies for the duration of the simulation, demonstrating persistence.
\begin{figure}
\centering
\includegraphics[width=1.0\columnwidth]{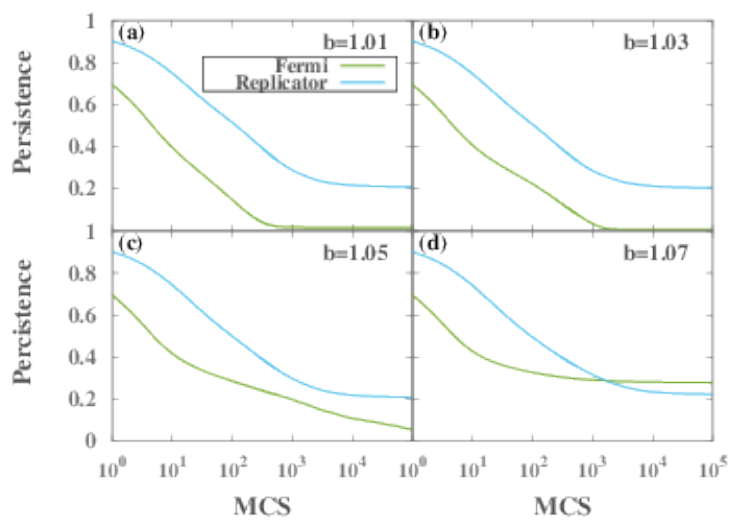} 
\caption{Temporal evolution of the persistence (percentage of individuals that maintain their initial strategies unchanged) for Fermi (green) and Replicator (blue) rules with $\rho=0.75$ and different values of temptation $b$: (a) $1.01$, (b) $1.03$, (c) $1.05$, and (d) $1.07$. Note that for the Replicator dynamics, the asymptotic fraction of unchanging sites is finite and almost independent of $b$.} 
\label{persistence} 
\end{figure}

To quantify persistence, in Fig.~\ref{persistence}, we present the percentage of individuals who keep their initial strategy unchanged throughout the simulation. We used $ \rho = 0.75 $, as in the previous figures. At least $ 20 \% $ of the agents remain persistent even after the system achieves equilibrium for REP (blue) for the temptation values  $ b = 1.01, 1.03 $ and $ 1.05 $ shown in Figs.~\ref{persistence} (a), (b), and (c), respectively. On the other hand, for the Fermi  rule (green), almost no agent maintained its initial strategy throughout the simulation. These results support the strong dependence on the initial conditions previously suggested for the REP rule. In Fig.~\ref{persistence}(d), the agents in the Fermi simulation are more persistent than in REP; however,  the lattice is quickly dominated by defectors ($ f_c \to 0 $, Fig.~\ref{fermi}) for this value of temptation ($b=1.07$), and some individuals never have the opportunity to change their strategies.

\begin{figure}
\includegraphics[angle=0, width=08cm]{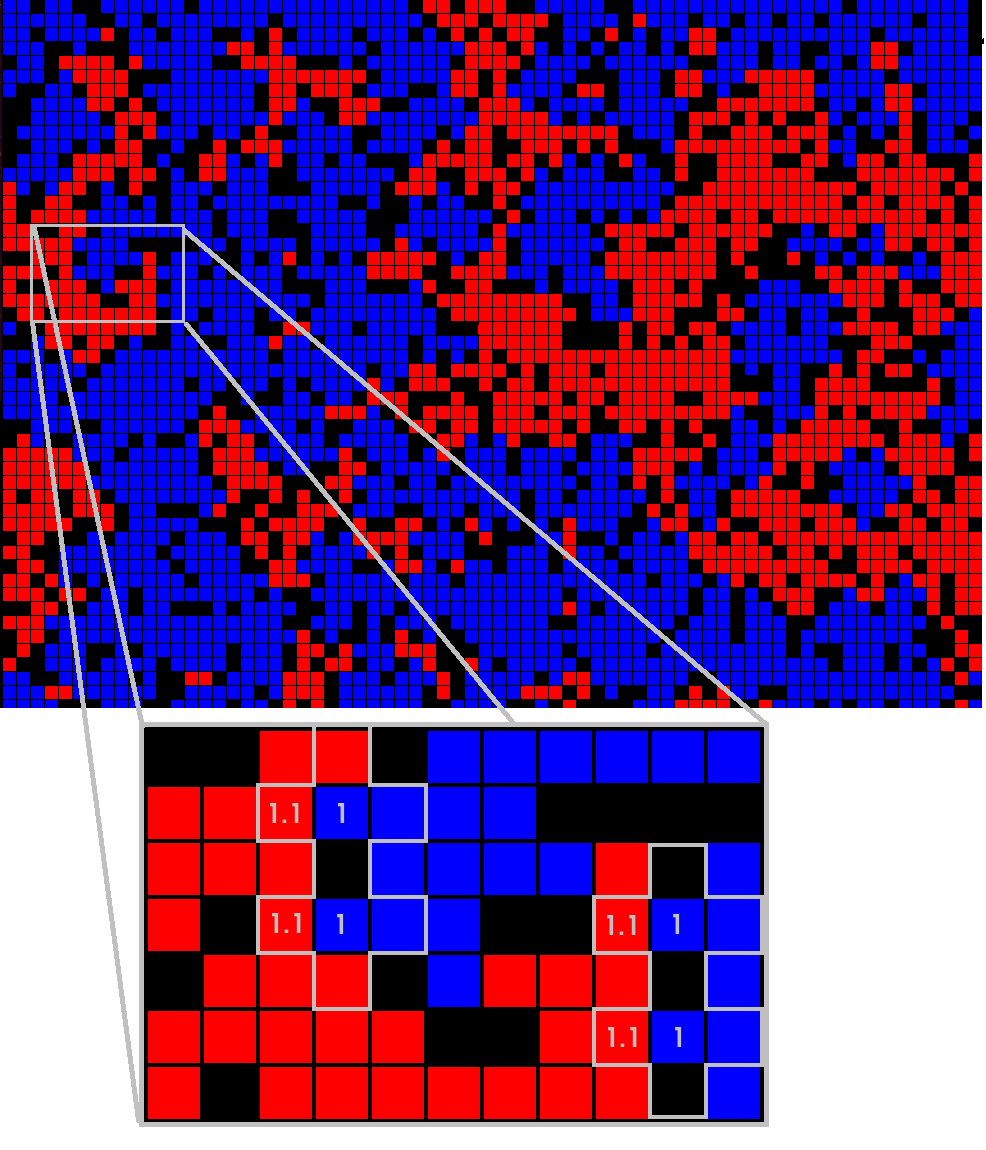}
\caption{(Color online) Snapshots of the asymptotic state of the system under  Replicator dynamics after $10^4$ MCS for $\rho=0.75$, $b=1.1$, and $L=70$. Cooperators are displayed in blue (dark gray), defectors in red (light gray) and holes in black. The zoomed in region shows the highlighted individual's payoff. Note that defectors having a higher payoff than their cooperator neighbors prevent invasion of defector clusters in the case of the Replicator rule.}
\label{snap_rpl}
\end{figure}

As the dynamics under the REP update rule strongly depends on the initial conditions, we may conclude that this rule does not allow the system to efficiently explore its entire configuration space, despite being probabilistic. To address this apparent contradiction, we will examine the system at the microscopic level. Figure~\ref{snap_rpl} presents a snapshot of the system's asymptotic state under the unperturbed REP update rule. Defectors and cooperators appear to be arranged in specific structures, with holes often present at the edges of defector clusters, creating boundaries between strategies.

Since the REP rule does not allow irrationality (as defined in the methods section), the zoomed-in region in Fig.~\ref{snap_rpl} shows arrangements where defectors will never change their strategy due to having higher payoffs than their cooperator neighbors, as seen in the highlighted interactions. In these cases, the cooperators have only one cooperating neighbor, while the others are holes or defectors, resulting in a payoff equal to $ R $. In contrast, their defector neighbor has a payoff of at least $ T = b$ and since $ T > R $, the defector will always maintain its strategy, preventing  cooperators from disseminating their strategy. This type of configuration may be the primary reason the system cannot explore the entire configuration space, obstructing waves of strategy changes from spreading efficiently throughout the lattice. As a result, the system becomes trapped in a frozen state. In short, the holes inhibit the flow of information by pinning the system. A similar phenomenon was previously described~\cite{VaAr01}, where pinning led to persistent sites that did not change strategy and to a cooperative peak at $\rho^*\approx 0.95$ for the ``choose the best'' update rule.

\begin{figure}
\begin{center} 
\includegraphics[angle = 270, width = 8cm] {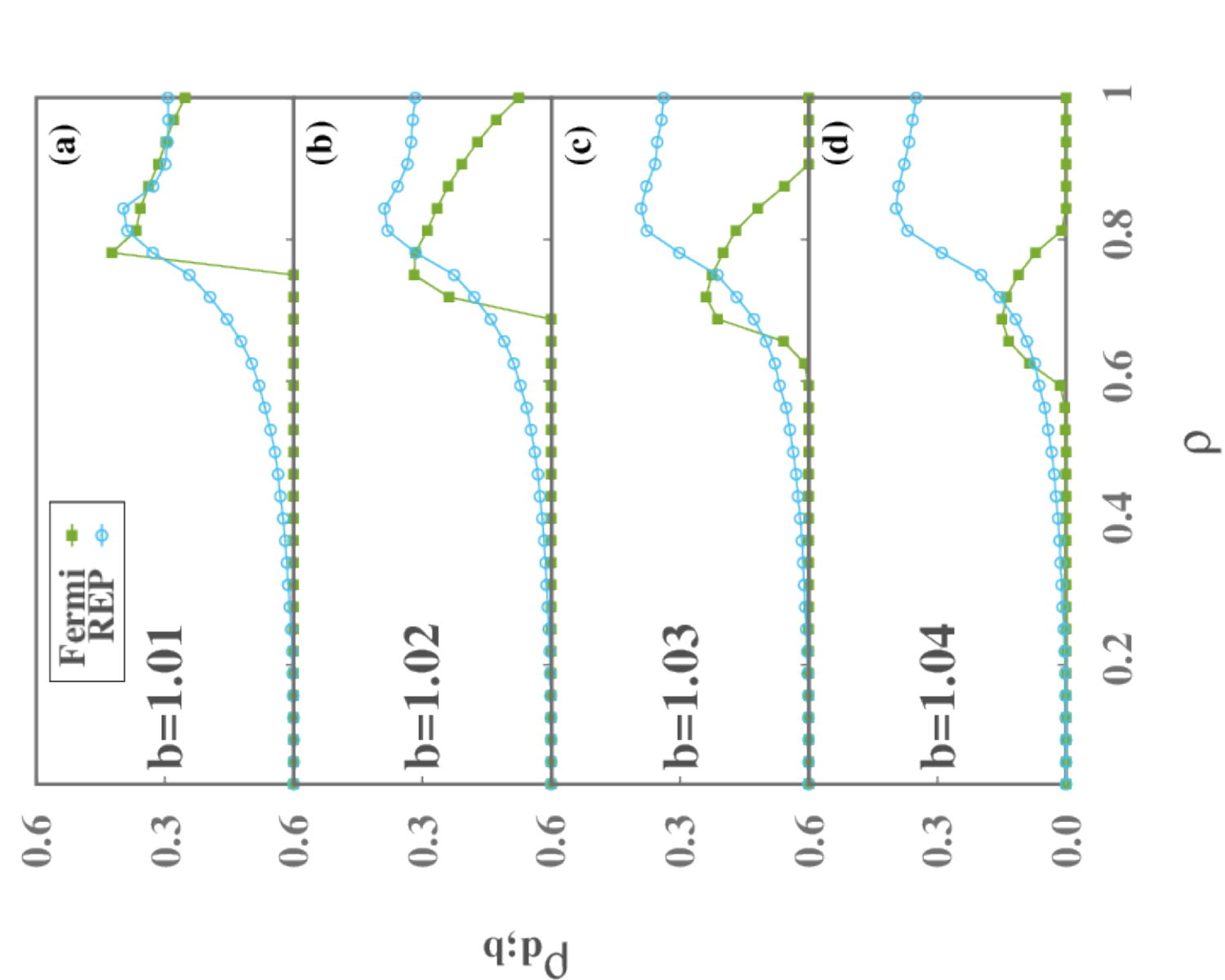} 
\end{center} 
\caption{Asymptotic fraction of defectors with payoff equal to $ T = b $ (only one cooperating neighbor) as a function of the population density $ \rho $ for  Fermi (green) and the Replicator (blue) rules for different values of the temptation $ b $. As in the case of persistence, it is practically independent of $b$ for the Replicator dynamics and, in this case, the location of the maximum coincides with the peak in cooperation. Note that in this case the maximum is located near the site percolation threshold $\rho_p\approx0.59$, as cooperation is driven to extinction.} 
\label{payoff} 
\end{figure}

To quantify the presence of this specific configuration, we measure the density of defectors with only one cooperator neighbor, which we will refer to as the $1D-1C$ case, among all defector-cooperator pairs. The density of this configuration can be expressed as
\begin{equation}
\rho_{(D,1b)}=\frac{N_{(D,1b)}}{\sum^{4}_{i=0}N_{(D,ib)}}
\end{equation}
where $\rho_{(D,1b)}$ is the asymptotic density of defectors with only one cooperator neighbor, and $N_{(D,1b)}$ represents the number of such defectors. In Fig.~\ref{payoff}, we present the behavior of $\rho_{(D,1b)}$ for the $1D-1C$ case as $\rho$ varies and compare both dynamical rules across different values of $b$. In the Fermi case (green lines), $\rho_{(D,1b)}$ rapidly decreases  at densities below the optimal density ($\rho \lesssim \rho^*$), while at higher $\rho$ values, the outcome becomes highly dependent on $b$. On the other hand, for REP dynamics (blue lines), $\rho_{(D,1b)}$ only vanishes at very low values of $\rho$, and its general behavior remains similar regardless of variations in $b$. The maintenance of the $1D$-$1C$ configuration for $\rho \lesssim \rho^*$ in REP highlights an important difference in the distribution of strategies compared to the Fermi rule. Additionally, the fact that REP curve maintains a similar behavior as   $b$ varies and has relatively fixed maximum whose location coincides with the peak in cooperation in Fig.~\ref{rep} suggests that this configuration is responsible for the pinning effect that creates this maximum.

The $1D$-$1C$ configuration in REP dynamics causes some individuals to become persistent due to the prohibition of irrationality. By introducing perturbations, cooperators can invade these structures, destroying the frozen state. As a result, we observe the expected behavior for stochastic dynamics—a peak near the percolation threshold, as shown in Fig.~\ref{perturb}. Interestingly, a similar effect can be achieved by introducing self-interaction, as seen in Fig.~\ref{densc1}, which allows cooperators to gain higher payoffs and invade defector clusters. However, for $\rho > 0.85$, the results of the perturbed REP simulation remain stable, and this analysis no longer applies. This stability is easily understood when we recognize that at high lattice densities, the number of $1D$-$1C$ configurations is too low to pin clusters and block the flow of strategies; thus, the system will return to its original configuration after perturbation, unless dominated by a single strategy during the process.

\begin{figure}
\includegraphics[angle=270, width=1\columnwidth]{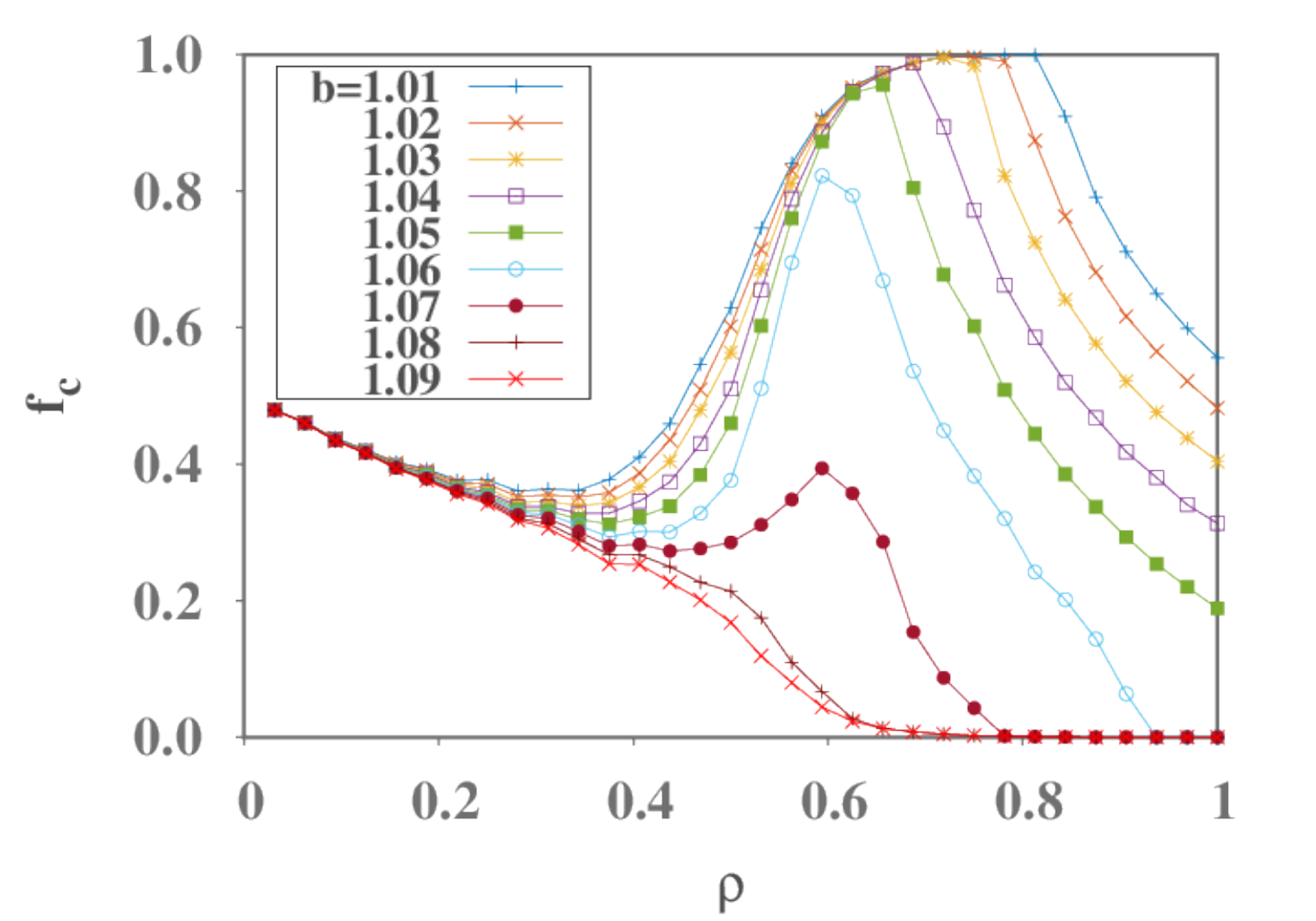}
\caption{Asymptotic fraction of cooperators $f_c$ as a function of the population density $\rho$ for different values of the temptation to defect $b$ for the Prisoner's Dilemma on the square lattice under synchronous updating. The individuals use the Replicator rule with a small fixed probability $\beta=0.08$ to change their strategy if the REP transition does not lead to an actual change, guaranteeing an uncertainty to strategy adoption.}
\label{ruidok}
\end{figure}

To conclude our analysis, we also tested the REP rule by introducing forced irrational decisions for all individuals through noise. Agents updated their strategy using the REP rule, but if they chose not to change, there was still a small probability ($\beta=0.08$) that they would switch strategies. As shown in Fig.~\ref{ruidok}, when all individuals are exposed to noise, the cooperative peak appears near the percolation threshold, as expected. A similar result is presented in~\cite{WaSzPe12} using a stochastic version of the  ``choose the best'' update rule.

Interestingly, the spatial structures identified in Fig.~\ref{snap_rpl} lead to a locally emergent deterministic behavior, even in a probabilistic system. However, self-interaction or perturbation can reintroduce the probabilistic bias, leading to the reappearance of the relationship between $\rho^*$ and $\rho_p$.

\section{Conclusions}
Previous studies have revealed that spatial structures play a crucial role in the maintenance of cooperation. The percolation threshold ($\rho_p$) is often linked to the optimum population density ($ \rho^* $) for cooperation, but only when strategy adoption involves some level of uncertainty. Without uncertainty, information does not spread efficiently, and the outcome depends heavily on initial conditions. To identify the quantitative factors responsible for a density near to $\rho_p$ that is both diluted and sufficiently connected to allow  the global optimum, we used the stochastic Fermi update rule and analyzed the distribution of agent clusters over time.

We argue that for densities $\rho > \rho_p$, the final number of cooperators and defectors is determined by a large percolating cluster that occupies nearly the entire lattice. In contrast, for $\rho < \rho_p$, the lattice becomes fragmented, with each fragment contributing independently to the total fraction of cooperators. When $\rho \approx \rho_p$, the extensive network connectivity allows strategies to spread efficiently. For this reason, at low values of $b$, which favor cooperation, $\rho^*$ reaches higher densities. Similarly, at high $b$, which favors defection, cooperation decreases rapidly for $\rho \gtrsim \rho_p$, and we observe $\rho^* \approx \rho_p$.

Unlike in Fermi dynamics, the REP update rule, despite being probabilistic, does not exhibit the expected relationship between $ \rho^* $ and $\rho_p$. We conclude that this unexpected behavior results from structures formed by holes and defectors, which prevent some defectors from being exposed to the noise associated with this update rule. As a result, deterministic patterns emerge even with a probabilistic rule. This occurs because defectors benefit from the presence of holes, a consequence of the Prisoner's Dilemma condition $ T = b > R = 1 $ and the fact that REP does not allow irrationality, meaning agents do not change their strategy when they have a higher payoff. However, since these structures require only a small number of holes to form, we observe a transition near $ \rho^* \approx 0.85$ for the REP dynamics. Above this threshold, there is no pinning of clusters and strategies change freely. On the other hand, for $ \rho < 0.85 $, the patterns that protect defectors from fluctuations emerge, making some agents persistent and preventing information from spreading efficiently. By introducing self-interaction or adding uncertainty to the Replicator rule, we achieve the expected relationship between the cooperative peak $ \rho^* $ and the percolation threshold $ \rho_p$.

Our results demonstrate that changes in lattice topology can alter the probabilistic nature of a stochastic update rule, creating deterministic regions. In other words, since the update rule governs how agents perceive and respond to their environment, changes in lattice topology can influence the outcomes of the update rule itself, altering how agents interact with their environment. We hope these findings contribute to a better understanding of how diluted lattices affect cooperation in Game Theory.

\begin{acknowledgments}
We thank Marco Antonio Amaral and Jefferson Arenzon for the their precious contributions. H.C.M.F. and M.H.V. acknowledge the financial support from the National Council for Scientific and Technological Development – CNPq (proc. 402487/2023-0). The simulations were conducted using the \href{https://pnipe.mcti.gov.br/laboratory/19775}{VD Lab} 
 cluster infrastructure at IF-UFRGS. F. R. Leivas acknowledges the financial agency CAPES (Coordenação de Aperfeiçoamento de Pessoal de Nível Superior) for providing her PhD scholarship.
\end{acknowledgments}


\bibliographystyle{apsrev4-1}
%

\end{document}